\documentclass[let,twocolumn]{jpsj3}
\usepackage{txfonts}
\usepackage{enumitem}
\usepackage{bm}

\usepackage[dvipdfmx]{graphicx}
\usepackage{tikz}
\usepackage{color} 

\usepackage{algorithm}
\usepackage{algpseudocode}
\usepackage{amsmath}
\usepackage{amssymb}
\usepackage{float}

\title{Population Annealing as a Discrete-Time Schrödinger Bridge}

\author{Masayuki Ohzeki}
\inst{Graduate School of Information Sciences, Tohoku University, Sendai 980-8579, Japan \\
Department of Physics, Tokyo Institute of Technology, Meguro, Tokyo 152-8551, Japan\\
Research and Education Institute for Semiconductors and Informatics, Kumamoto University, Kumamoto 860-8555, Japan\\
Sigma-i Co., Ltd., Minato, Tokyo 108-0075, Japan}

\abst{
We present a theoretical framework that reinterprets Population Annealing (PA) through the lens of the discrete-time Schrödinger Bridge (SB) problem.
We demonstrate that the heuristic reweighting step in PA is derived by analytically solving the Schrödinger system without iterative computation via instantaneous projection.
In addition, we identify the thermodynamic work as the optimal control potential that solves the global variational problem on path space.
This perspective unifies non-equilibrium thermodynamics with the geometric framework of optimal transport, interpreting the Jarzynski equality as a consistency condition within the Donsker-Varadhan variational principle, and elucidates the thermodynamic optimality of PA.
}

\begin{document}
\maketitle

\section{Introduction}
Simulating equilibrium states of complex systems characterized by rough energy landscapes is a central challenge in statistical physics.
Standard Markov Chain Monte Carlo (MCMC) methods often suffer from slow convergence due to trapping in local minima.
To overcome this ergodicity breaking, various generalized ensemble methods have been proposed, such as Replica Exchange Monte Carlo (Parallel Tempering)\cite{Hukushima1996} and Simulated Tempering\cite{marinari1992}.
Among these, Population Annealing (PA)\cite{Hukushima2003, Machta2010} has recently emerged as a particularly powerful approach.
By evolving a population of replicas under a changing temperature schedule, PA enables efficient parallel sampling and, crucially, provides a direct estimator of free energy differences via the measurement of non-equilibrium work based on the Jarzynski equality, which often supports an efficient computing in stochastic dynamics and quantum gate model as well \cite{Katsuda2011,Ohzeki2010}.

In a separate development within applied mathematics and machine learning, the Schrödinger Bridge (SB) problem\cite{schrodinger1931, leonard2014} has gained renewed attention.Originally posed by Schrödinger to find the most likely evolution of a probability density between two fixed marginals, SB is now recognized as a formulation of entropy-regularized optimal transport\cite{peyre2019}.
It has become a foundational framework for modern generative modeling, particularly in diffusion models\cite{song2021, debortoli2021, Kaba2025} and stochastic control\cite{chen2021}, where iterative algorithms are typically employed to learn the optimal control force.

In this context, algorithmic connections between Sequential Monte Carlo (SMC) methods and SB have also been explored.
For instance, Bernton et al.\cite{bernton2019} proposed improving SMC proposals using iterative SB formulations (Iterative Proportional Fitting), and Liu et al.\cite{liu2025} developed efficient adjoint-based samplers.
These works primarily focus on algorithmic refinements to enhance sampling efficiency through iterative optimization and learning.

While these studies position SMC methods as algorithmic components within iterative SB solvers, Population Annealing, which is one of the SMC methods, is used in physics as a standalone, non-iterative sampler.
However, the standard formulation of the SB problem typically requires iterative procedures, such as the Sinkhorn algorithm, to strictly satisfy boundary constraints at both the initial and final distributions.
It is therefore not immediately obvious how PA, which operates purely sequentially without such global iterations, relates to the rigorous optimal transport framework.
Does PA merely approximate the transport, or does it constitute a valid solution to the SB problem under specific conditions?
The physical geometry underlying this efficiency—specifically, how PA bypasses the need for iteration via thermodynamic principles—has not been fully explicated.

In this Letter, we reorganize the theoretical understanding of PA by explicitly mapping it to the discrete-time SB framework.
Unlike previous works that focus on iterative refinement, we demonstrate that the heuristic reweighting step in PA corresponds exactly to the optimal control term required to solve the SB problem analytically in the instantaneous projection limit.
Furthermore, we extend the discussion from a single step to the entire annealing schedule, proving that minimizing the local transport cost step-by-step is equivalent to solving the global optimal transport problem.
This formulation unifies the physical concept of the Jarzynski equality with the geometric perspective of optimal transport, suggesting that PA is a natural solver for the global Donsker-Varadhan variational problem on path space.

\section{Population Annealing and Free Energy Estimation}
We consider a system with state $x \in \mathcal{X}$ and energy function $E(x)$.
Our goal is to sample from the Boltzmann distribution $\pi_k(x) \propto e^{-\beta_k E(x)}$ across a sequence of inverse temperatures $\beta_0 < \beta_1 < \dots < \beta_K$.

In the PA algorithm, at step $k$, the temperature is lowered from $\beta_k$ to $\beta_{k+1}$.
PA adjusts the population by calculating a weight $w_k^{(i)}$ for each replica $i$:
\begin{equation}
w_k(x) = \frac{e^{-\beta_{k+1} E(x)}}{e^{-\beta_k E(x)}} = e^{-(\beta_{k+1} - \beta_k) E(x)}.
\end{equation}
The population is then resampled proportional to these weights, followed by an MCMC transition (mutation).
This process has a direct thermodynamic interpretation where $\ln w_k(x) = -\Delta \beta_k E(x)$ corresponds to the thermodynamic work $W_k$.
The average of these weights estimates the free energy difference via the Jarzynski equality\cite{jarzynski1997}: $\langle e^{-W_k} \rangle_{\pi_k} = e^{-\Delta F_k}$.

\section{The Discrete-Time Schrödinger Bridge}
To formalize PA within the optimal transport framework, we consider the Schrödinger Bridge (SB) problem on the path space $X_{0:K} = (x_0, \dots, x_K)$.
Let $\mathcal{Q}$ be the probability measure of the reference Markov process (uncontrolled MCMC) with joint distribution $\mathcal{Q}(x_{0:K}) = \pi_0(x_0) \prod_{k} M_k(x_{k+1}|x_k)$, where $\pi_t(x)$ is the Gibbs-Boltzmann distribution as a steady state.
On the other hand, we define the modified path probability as $\mathcal{P}(x_{0:K}) = \pi_0(x_0) \prod_{k} M'_k(x_{k+1}|x_k)$, which is given by a different kernel $M'_k(x_{k+1}|x_k)$.

The global SB problem seeks a path measure $\mathcal{P}$ that minimizes the KL divergence from $\mathcal{Q}$:
\begin{equation}
    \min_{\mathcal{P}} D_{KL}(\mathcal{P} \| \mathcal{Q}).
\end{equation}
The nature of the solution depends critically on the constraints imposed on $\mathcal{P}$.

The standard formulation of the SB problem imposes constraints only on the initial and final distributions:
\begin{equation}
    P_0(x_0) = \pi_0(x_0), \quad P_K(x_K) = \pi_K(x_K).
\end{equation}
Here $P_t(x_t)$ is an instantaneous probability distribution.
In this case, the optimal path measure $\mathcal{P}^*$ is known to be a Markov bridge.
The KL divergence is reduced to the following form:
\begin{align}
    D_{KL}(\mathcal{P} \| \mathcal{Q}) &= \sum_{k=0}^{K-1} \sum_{x_{k+1},x_k} P_k(x_k) M'_k(x_{k+1}|x_k) \log \frac{M'_k(x_{k+1}|x_k) }{M_k(x_{k+1}|x_k)}.
\end{align}
In general, the KL divergence between path measures decomposes into the expectation of the sum of conditional KL divergences given the full history $x_{0:k}$.
However, in this case, the stochastic dynamics are governed by a Markov process.

The optimal ``joint distribution" $P_k^*(x_k,x_{k+1})$ at adjacent time steps $k, k+1$ is given by the symmetric modification of the reference measure:
\begin{equation}
    P_k^*(x_k, x_{k+1}) = \varphi_k(x_k) Q_k(x_k, x_{k+1}) \psi_{k+1}(x_{k+1}), \label{eq:optimal_joint}
\end{equation}
where $Q_k(x_k, x_{k+1}) = \pi_k(x_k) M_k(x_{k+1}|x_k)$ is the reference joint distribution.
$\varphi_k$ and $\psi_k$ are space-time potentials.
Notice that $P^*_k(x_k,x_{k+1})$ is not necessarily normalized.
The following manipulation does not require kernel normalization.

In this case, the optimal path measure $\mathcal{P}^*$ is known to be a Markov bridge given by the multiplicative modification of the reference measure.
The space-time potentials $\varphi_k$ and $\psi_k$ must satisfy the Schrödinger system, which consists of the forward and backward recurrence relations:
\begin{align}
\varphi_{k+1}(y) = \sum_{x} \varphi_k(x) Q_k(x, y), \label{eq:sb_forward} \\
\psi_k(x) = \sum_{y} Q_k(x, y) \psi_{k+1}(y).
\label{eq:sb_backward}
\end{align}
These equations describe the propagation of boundary information: $\varphi_k$ carries the initial constraint $\pi_0$ forward, while $\psi_k$ propagates the final constraint $\pi_K$ backward.
It implies that the marginal distribution at any time step $k$ is given by the pointwise product of the two potentials:
\begin{equation}
P_k(x) = \varphi_k(x) \psi_k(x).
\end{equation}
This symmetric factorization highlights the role of $\varphi_k$ as the forward message from the initial state and $\psi_k$ as the backward message from the final constraint.
The solution is determined by the boundary conditions:
\begin{align}
\varphi_0(x) \psi_0(x) = \pi_0(x), \quad \varphi_K(x) \psi_K(x) = \pi_K(x).
\end{align}
Since the equations are coupled globally across time $0 \le k \le K$, one typically employs the iterative Sinkhorn algorithm to find the potentials that satisfy both boundaries simultaneously.

However, this iterative procedure is computationally expensive, especially for high-dimensional systems, as it requires storing and updating the entire history of potentials.

\section{Bridging PA and SB}

To circumvent this computational burden, we introduce a stronger set of constraints to simplify the problem structure.
Instead of fixing only the endpoints, we enforce the equilibrium distribution $\pi_k$ as a marginal constraint at every intermediate time step:
\begin{equation}
P_k(x_k) = \pi_k(x_k) \quad \text{for all } k=0, \dots, K.
\end{equation}
By pinning the distribution to the equilibrium target $\pi_k$ at each moment, we break the global temporal coupling.
This transforms the global optimization problem into a sequence of local, independent transport problems between adjacent steps $\pi_k$ and $\pi_{k+1}$, which allows for a direct, non-iterative solution.

Moreover, this formulation aligns naturally with the physical reality of annealing protocols.
In such experiments or simulations, an external operator arbitrarily manipulates control parameters (such as the inverse temperature $\beta_k$) according to a predetermined schedule.
This operation explicitly defines a sequence of target equilibrium distributions $\pi_k$ that the system is intended to track at each step, rather than merely imposing a constraint on the final state.

The potentials $\varphi_k$ and $\psi_{k+1}$ are determined by the marginal constraints imposed by the local minimization of the KL divergence.
Marginalizing the optimal joint distribution $P_k^*(x_k, x_{k+1})$ with respect to $x_{k+1}$ and $x_k$ must recover the fixed distributions $\pi_k(x_k)$ and $\pi_{k+1}(x_{k+1})$, respectively.
This yields the following coupled equations:
\begin{align}
\varphi_k(x_k) = \frac{\pi_k(x_k)}{\sum_{x_{k+1}} Q_k(x_k, x_{k+1}) \psi_{k+1}(x_{k+1})}, \label{eq:sinkhorn_phi} \\
\psi_{k+1}(x_{k+1}) = \frac{\pi_{k+1}(x_{k+1})}{\sum_{x_k} \varphi_k(x_k) Q_k(x_k, x_{k+1})}.
\label{eq:sinkhorn_psi}
\end{align}
One can recognize that this system is overdetermined if we attempt to solve both equations analytically for a single step.
However, we can exploit the freedom to fix one potential.
If we substitute $\psi_{k+1}(x_{k+1})=1$ into Eq.~(\ref{eq:sinkhorn_phi}), we obtain the trivial solution $\varphi_k(x_k)=1$ (since $\sum_{x_{k+1}} Q_k = \pi_k$), which implies no distributional change.
In contrast, if we postulate $\varphi_k(x_k)=1$ in Eq.~(\ref{eq:sinkhorn_psi}), we can calculate $\psi_{k+1}(x_{k+1})$ analytically.
This choice allows us to strictly satisfy the target marginal condition at each time step.
The solution obtained from this latter path corresponds exactly to the Population Annealing algorithm.

The ``forward equation" in the Sinkhorn algorithm yields the analytical solution for the potential:
\begin{equation}
\psi_{k+1}(y) = \frac{\pi_{k+1}(y)}{\pi_k(y)}.
\end{equation}
This result signifies that since the mutation kernel $M_k$ does not alter the equilibrium distribution $\pi_k$, the entire distributional shift to $\pi_{k+1}$ must be induced solely by the potential $\psi_{k+1}$.
In PA, this is implemented by the resampling step.
Substituting this into the joint distribution $P^*$:
\begin{equation}
P^*(x_{k+1}, x_k) = \frac{\pi_{k+1}(x_{k+1})}{\pi_k(x_{k+1})} M_k(x_{k+1}|x_k) \pi_k(x_k) .
\end{equation}
Since we established that the global cost function is the sum of local costs, applying this optimal control at every step ensures global optimality across the entire schedule $0 \to K$.
Substituting the PA optimal solution into the cost function, the log-likelihood ratio simplifies significantly:
\begin{equation}
D_{KL}(\mathcal{P}\|\mathcal{Q}) =\sum_{k=0}^{K-1} D_{KL}(\pi_{k+1} \| \pi_k).
\end{equation}
Substituting the definition of the canonical distributions, we find that this cost function is exactly equal to the thermodynamic dissipated work.
Let us define the generalized work performed during the parameter update from $\beta_k$ to $\beta_{k+1}$ as $W_k(x) = (\beta_{k+1} - \beta_k)E(x)$ and the dimensionless free energy as $\phi_k = -\ln Z_k$.
The KL divergence can be expanded as:
\begin{eqnarray}\nonumber
D_{KL}(\pi_{k+1} \| \pi_k) 
&=& \Delta \phi - \langle W_k \rangle_{\pi_{k+1}}.
\end{eqnarray}
This quantity represents the entropy production or the dissipated work defined by the second law of thermodynamics: the difference between the free energy change and the average work.
Thus, minimizing the transport cost in PA is rigorously equivalent to minimizing the strictly defined thermodynamic dissipation.
Beyond the known connection to the Jarzynski equality, our work reveals the fundamental variational structure of the algorithm.
We assert that the PA algorithm is obtained by minimizing the Schrödinger Bridge objective subject to strict intermediate constraints.
This formulation reveals that the mathematical cost function is strictly equivalent to the physical entropy production defined in thermodynamics.

Furthermore, if we consider the limit of small step sizes $\Delta \beta =\beta_{k+1}-\beta_k \to 0$, this exact quantity can be approximated by the second-order expansion involving the Fisher information (energy variance):
\begin{equation}
D_{KL}(\pi_{k+1} \| \pi_k) \approx \frac{1}{2} \mathrm{Var}(E)_{\beta_k} (\Delta \beta_k)^2.
\end{equation}
Since the magnitude of the resultant cost quantifies the inherent difficulty of the distributional transport, minimizing it corresponds to realizing control with minimal thermodynamic effort.
This metric serves as a fundamental criterion for optimizing the annealing schedule: step sizes should be reduced in regions where the cost (energy variance) is high to maintain ``effortless" control, and can be increased where the cost is low.

Interestingly, the criterion of constant thermodynamic speed is physically equivalent to the optimal scheduling condition in Replica Exchange Monte Carlo (EMC, also known as Parallel Tempering) \cite{Hukushima1996}, where the temperature intervals are chosen to keep the acceptance probabilities of replica swaps constant.
Both algorithms ultimately demand the same geometric optimization: ensuring sufficient overlap between adjacent energy distributions to facilitate efficient transport (swapping in EMC, or resampling in PA).
This geometric equivalence strongly suggests that the transition rules of EMC can also be formally derived as an optimal solution within the discrete-time Schr\"odinger Bridge framework.
However, a rigorous mapping of the EMC algorithm onto the SB framework requires a detailed treatment of the joint distributions and their corresponding potentials.
Therefore, we leave this comprehensive discussion for a separate future publication.

We now address the temporal structure of the control.
In standard SB, the solution is often a ``dynamical" control (modifying $M_k$ during transition).
However, the discrete-time setting allows for an alternative: instantaneous control.
In PA, the parameter switch $\beta_k \to \beta_{k+1}$ creates a discrepancy.
The likelihood ratio required to bridge this gap is related to the resampling weight $w_k(x)$ defined in Eq.~(1) and the partition functions:
\begin{equation}
\frac{\pi_{k+1}(x)}{\pi_k(x)} = \frac{e^{-\beta_{k+1} E(x)}/Z_{k+1}}{e^{-\beta_k E(x)}/Z_k} = w_k(x)\frac{Z_k}{Z_{k+1}}.
\label{eq:ratio_w}
\end{equation} 
By resampling with $w_k(x)$, PA actively utilizes this work to ``bridge the gap" immediately.
It effectively resets the empirical measure to $\pi_{k+1}$ before the unmodified kernel $M_k$ is applied.

In the language of the Schrödinger system, this implies that the iterative Sinkhorn procedure is rendered unnecessary.
Because the control is applied as an instantaneous projection (resampling) and the reference kernel preserves the source distribution, the optimal potential $\psi$ is analytically determined by the Boltzmann factor ratio.

Thus, the resampling step in PA is an exact algorithmic implementation of the optimal control potentials.
The ``work" $W_k$ corresponds directly to the log-potential required to perform the optimal control, and the sequence of these operations constitutes the optimal transport path for the entire annealing process.

\section{Geometric Interpretation on Path Space}
The identification of PA as a fully constrained global Schrödinger Bridge solver offers deeper insight into the mathematical nature of non-equilibrium work relations.
We consider the probability measures on the entire path space $X_{0:K} = (x_0, \dots, x_K)$.

Let $\mathcal{Q}$ be the reference path measure.
The optimal transport measure $\mathcal{P}^*$ constructed by PA is related to $\mathcal{Q}$ via a change of measure.
The ratio of their densities is the global Radon-Nikodym derivative.
Using Eq.~(\ref{eq:ratio_w}):
\begin{equation}
\frac{d\mathcal{P}^*}{d\mathcal{Q}}(\mathbf{x}) = \prod_{k=0}^{K-1} \frac{\pi_{k+1}(x_{k+1})}{\pi_k(x_{k+1})} \approx \prod_{k=0}^{K-1} w_k(x_k) \frac{Z_k}{Z_{k+1}}.
\end{equation}
The global Jarzynski equality arises as the normalization condition for this global measure transformation.
Taking the expectation over the reference path ensemble $\mathcal{Q}$:
\begin{equation}
    \langle e^{-\mathcal{W}} \rangle_{\mathcal{Q}} = \frac{Z_K}{Z_0} \langle \frac{d\mathcal{P}^*}{d\mathcal{Q}} \rangle_{\mathcal{Q}} = \frac{Z_K}{Z_0} = e^{-\Delta F_{\text{total}}}.
\end{equation}
Thus, in the global SB framework, the Jarzynski equality serves as the geometric consistency condition.

In addition, the optimality of PA is grounded in the Donsker-Varadhan variational formula \cite{donsker1975}.

By identifying the cumulative work $\mathcal{W}$ as the optimal control potential, the free energy relation is recast as a variational principle on the path space:
\begin{equation}
-\Delta F_{\text{total}} = \ln \langle e^{-\mathcal{W}} \rangle_{\mathcal{Q}} = \sup_{\mathcal{P}} \left\{ -\langle \mathcal{W} \rangle_{\mathcal{P}} - D_{KL}(\mathcal{P} \| \mathcal{Q}) \right\}.
\end{equation}
This objective maximizes energetic efficiency (negative work) penalized by the information-theoretic cost of deviating from the reference dynamics, effectively tightening the standard Second Law bound $\langle \mathcal{W} \rangle \ge \Delta F_{\text{total}}$.
Since the SB problem corresponds to the minimization of the KL divergence, it represents the mathematical dual to this maximization.
PA, by reweighting trajectories proportional to $e^{-\mathcal{W}}$, empirically constructs the unique optimal path measure $\mathcal{P}^*$ that attains this supremum, thereby solving the global variational problem for finite-time transport.
In this light, the Jarzynski equality serves as the geometric consistency condition, ensuring that the optimal path measure is properly normalized to bridge the initial and final distributions.

\section{Conclusion}
In this Letter, we have established that Population Annealing is not merely a heuristic algorithm but a rigorous solver for the discrete-time Schrödinger Bridge problem.
While recent machine learning literature has focused on iterative refinements of such samplers, our reformulation clarifies the physical mechanism behind PA's efficiency.
We demonstrated that the resampling step constitutes an analytical solution to the Schrödinger system via static projection.
By explicitly contrasting the standard endpoint-constrained SB problem, which necessitates iterative Sinkhorn computations, with the sequentially constrained PA formulation, we clarified why PA admits a non-iterative analytical solution.
This analysis reveals that the resampling step acts as an instantaneous static projection, providing a globally optimal transport trajectory on the path space.

The significant insight gained from this perspective is the identification of the cumulative thermodynamic work as the exact potential required for optimal transport.
This framework offers a dual unification: physically, it reinterprets the Jarzynski equality as a geometric consistency condition within the Donsker-Varadhan variational principle.
From an algorithmic perspective, it bridges statistical physics and machine learning by characterizing PA as a training-free, non-iterative strategy for optimal transport.
This insight opens new avenues for efficient sampling in energy-based generative modeling.

Finally, a promising future direction is to extend this framework to dynamics breaking detailed balance.
Since non-reversible processes are known to accelerate convergence\cite{suwa2010, sakai2013, ichiki2013, ohzeki2015, ohzeki2015conf}, utilizing optimal transport theory to design and reweight such non-equilibrium dynamics could lead to even more efficient sampling algorithms beyond the standard limitations of reversibility.

We acknowledge that the standard SB problem, which imposes constraints only at the endpoints, theoretically admits a broader class of solutions, including the PA trajectory.
However, the unique feature of PA is its greedy, sequential nature enabled by the intermediate constraints.
While the standard iterative SB approach can find the globally optimal path without these intermediate restrictions (potentially finding "tunneling" paths that standard annealing misses), PA provides a highly efficient "forward-pass" approximation that is exact when the system follows the prescribed temperature schedule.
Comparing these two regimes—sequential constrained transport versus global iterative transport—remains a promising avenue for future research.

\begin{acknowledgments}
This study was supported by the JSPS KAKENHI Grant No. 23H01432.
We received financial supports by programs for bridging the gap between R\&D and IDeal society (Society 5.0) and Generating Economic and social value (BRIDGE) and Cross-ministerial Strategic Innovation Promotion Program (SIP) from the Cabinet Office (No. 23836436).
\end{acknowledgments}

\bibliographystyle{jpsj}
\bibliography{main_v2}       

@article{Hukushima1996,
  title={Exchange Monte Carlo Method and Application to Spin Glass Simulations},
  author={Hukushima, Koji and Nemoto, Koji},
  journal={Journal of the Physical Society of Japan},
  volume={65},
  number={6},
  pages={1604--1608},
  year={1996},
  publisher={The Physical Society of Japan},
  doi={10.1143/JPSJ.65.1604}
}

@article{bernton2019,
  title={Schr{\"o}dinger Bridge Samplers},
  author={Bernton, Espen and Heng, Jeremy and Doucet, Arnaud and Jacob, Pierre E.},
  journal={arXiv preprint arXiv:1912.13170},
  year={2019}
}

@inproceedings{liu2025,
  title={Adjoint Schr{\"o}dinger Bridge Sampler},
  author={Liu, Guan-Horng and Choi, Jaemoo and Chen, Yongxin and Miller, Benjamin Kurt and Chen, Ricky T. Q.},
  booktitle={Proceedings of the 13th International Conference on Learning Representations (ICLR)},
  year={2025}
}

@article{Kaba2025,
  title={Schr{\"o}dinger bridge-type diffusion models as an extension of variational autoencoders},
  author={Kaba, Kentaro and Shimizu, Reo and Ohzeki, Masayuki and Sughiyama, Yuki},
  journal={Physical Review Research},
  volume={7},
  number={3},
  pages={033213},
  year={2025},
  publisher={APS},
  doi={10.1103/PhysRevResearch.7.033213}
}

@article{donsker1975,
  title={Asymptotic evaluation of certain Markov process expectations for large time. I},
  author={Donsker, Monroe D. and Varadhan, S. R. S.},
  journal={Communications on Pure and Applied Mathematics},
  volume={28},
  number={1},
  pages={1--47},
  year={1975},
  publisher={Wiley Online Library},
  doi={10.1002/cpa.3160280102}
}

@article{marinari1992,
  title={Simulated tempering: a new Monte Carlo scheme},
  author={Marinari, Enzo and Parisi, Giorgio},
  journal={Europhysics Letters},
  volume={19},
  number={6},
  pages={451--458},
  year={1992},
  publisher={IOP Publishing}
}

@inproceedings{Hukushima2003,
  title={Population annealing and its application to a spin glass},
  author={Hukushima, Koji and Iba, Yukito},
  booktitle={AIP Conference Proceedings},
  volume={690},
  number={1},
  pages={200--206},
  year={2003},
  organization={American Institute of Physics}
}

@article{Machta2010,
  title={Population annealing with weighted averages: A Monte Carlo method for rough free-energy landscapes},
  author={Machta, Jonathan},
  journal={Physical Review E},
  volume={82},
  number={2},
  pages={026704},
  year={2010},
  publisher={APS}
}

@article{Katsuda2011,
  title={Population Annealing for the Eigenvalue Problem},
  author={Katsuda, Hidetaka and Ohzeki, Masayuki},
  journal={Journal of the Physical Society of Japan},
  volume={80},
  number={4},
  pages={045003},
  year={2011},
  publisher={The Physical Society of Japan}
}

@article{Ohzeki2010,
  title={Nonequilibrium Equality for Free Energy Differences in Quantum Systems},
  author={Ohzeki, Masayuki},
  journal={Physical Review Letters},
  volume={105},
  number={5},
  pages={050401},
  year={2010},
  publisher={APS}
}

@article{schrodinger1931,
  title={{\"U}ber die Umkehrung der Naturgesetze},
  author={Schr{\"o}dinger, Erwin},
  journal={Sitzungsberichte der Preussischen Akademie der Wissenschaften. Physikalisch-mathematische Klasse},
  volume={144},
  pages={144--153},
  year={1931}
}

@article{leonard2014,
  title={A survey of the Schr{\"o}dinger problem and some of its connections with optimal transport},
  author={L{\'e}onard, Christian},
  journal={Discrete \& Continuous Dynamical Systems-A},
  volume={34},
  number={4},
  pages={1533--1574},
  year={2014}
}

@article{peyre2019,
  title={Computational optimal transport: With applications to data science},
  author={Peyr{\'e}, Gabriel and Cuturi, Marco},
  journal={Foundations and Trends{\textregistered} in Machine Learning},
  volume={11},
  number={5-6},
  pages={355--607},
  year={2019},
  publisher={Now Publishers, Inc.}
}

@inproceedings{song2021,
  title={Score-Based Generative Modeling through Stochastic Differential Equations},
  author={Song, Yang and Sohl-Dickstein, Jascha and Kingma, Diederik P and Kumar, Abhishek and Ermon, Stefano and Poole, Ben},
  booktitle={International Conference on Learning Representations},
  year={2021}
}

@article{debortoli2021,
  title={Diffusion Schr{\"o}dinger bridge with applications to score-based generative modeling},
  author={De Bortoli, Valentin and Thornton, James and Heng, Jeremy and Doucet, Arnaud},
  journal={Advances in Neural Information Processing Systems},
  volume={34},
  pages={17695--17709},
  year={2021}
}

@article{chen2021,
  title={Likelihood Training of Schr{\"o}dinger Bridge using Forward-Backward SDEs},
  author={Chen, Tianrong and Liu, Guan-Horng and Theodorou, Evangelos A},
  journal={arXiv preprint arXiv:2106.01357},
  year={2021}
}

@article{jarzynski1997,
  title={Nonequilibrium equality for free energy differences},
  author={Jarzynski, Christopher},
  journal={Physical Review Letters},
  volume={78},
  number={14},
  pages={2690},
  year={1997},
  publisher={APS}
}

@article{suwa2010,
  title={Markov chain Monte Carlo methods without detailed balance},
  author={Suwa, Hidemaro and Todo, Synge},
  journal={Physical Review Letters},
  volume={105},
  number={12},
  pages={120603},
  year={2010},
  publisher={APS}
}

@article{sakai2013,
  title={Population annealing simulation of spin glasses},
  author={Sakai, Yoshiki and Hukushima, Koji},
  journal={Journal of the Physical Society of Japan},
  volume={82},
  number={6},
  pages={064003},
  year={2013},
  publisher={The Physical Society of Japan}
}

@article{ichiki2013,
  title={Langevin dynamics derived from quantum walks},
  author={Ichiki, Akihisa and Ohzeki, Masayuki},
  journal={Physical Review E},
  volume={88},
  number={2},
  pages={020101},
  year={2013},
  publisher={APS}
}

@article{ohzeki2015,
  title={Langevin dynamics for the classical Ising model derived from quantum mechanics},
  author={Ohzeki, Masayuki and Ichiki, Akihisa},
  journal={Physical Review E},
  volume={92},
  number={1},
  pages={012105},
  year={2015},
  publisher={APS}
}

@article{ohzeki2015conf,
  title={Langevin dynamics derived from quantum mechanics},
  author={Ohzeki, Masayuki and Ichiki, Akihisa},
  journal={Journal of Physics: Conference Series},
  volume={638},
  number={1},
  pages={012003},
  year={2015},
  publisher={IOP Publishing}
}
\end{document}